# STATISTICS OF SMALL-$N$ SIMULATIONS

DOUGLAS C. HEGGIE

*Department of Mathematics and Statistics, University of Edinburgh, King's Buildings, Edinburgh EH9 3JZ, UK*

**Abstract.** We first review the reasons for carrying out statistical analysis of results from large numbers of $N$-body simulations, including a summary of previous work. Then we describe some results about the behaviour of $N$-body systems which have been acquired by this technique. Finally we concentrate on the problems associated with the scaling of $N$-body data with $N$, with particular regard to the dissolution of systems in a tidal field when mass is lost by stellar evolution.

## 1. Introduction

One of the frequent complaints against data from $N$-body simulations is that they are "noisy". If one examines Lagrangian radii for the evolution of a 250-body system, for example, the inner radii show evidence of core collapse but it is not easy to measure the rate of collapse, or even the time at which collapse ends. One way in which the data can be improved is by combining the results from many simulations. Experience shows that the statistical noise is reduced as one would naively expect, i.e. in proportion to the square root of the number of simulations. Thus combining results from 50 simulations improves the signal-to-noise by a factor of approximately 7. Every observer knows how much more science he could do with such an improvement, but theorists have been slower in seeing the benefits.

This paper first reviews this and other reasons for taking a statistical approach to the study of $N$-body simulations. After summarising some previous investigations carried out in this spirit, we turn to some recent work which highlights some of the advantages and pitfalls of this kind of work.

## 2. Advantages of a Statistical Approach

2.1. THEORETICAL ARGUMENTS

It has been known for a long time that the detailed results of $N$-body calculations, i.e. the positions and velocities of the individual stars, are wrong. The early work of Miller (1964, 1971) showed that the growth of errors is extraordinarily rapid, with consequences which were illustrated by Lecar (1968) and Hayli (1970). Recent studies (Kandrup & Smith 1991, 1992a,b) have extended this work to larger $N$ and given it a theoretical foundation (Goodman *et al.* 1993).

In the face of this observation, the justification for the validity of $N$-body simulations is that the *statistical* results of the computations are nevertheless correct (Aarseth & Lecar 1975). In statistics, however, it is usual to take many measurements, in a compromise between cost and precision. Since the only valid results of $N$-body simulations are statistical, the same approach should be used there.

The assertion that statistical results of $N$-body simulations are valid is an article of faith among practitioners, but it has been given some support recently by "shadowing lemmas" (Quinlan & Tremaine 1992; Hayes 1995). Roughly speaking, these show that there exists an exact $N$-body model which lies close to a given computed model, despite the numerical errors in the latter.

2.2. PRACTICAL ARGUMENTS

The strategy of computing many realisations of a single system, differing only in the random numbers used to create initial conditions, is ideally suited to implementation on parallel computers or even networks of workstations. Communications are a negligible part of the effort, and so virtually 100% efficiency is achieved. The main practical difficulties, in fact, stem from storing and analysing the rather large quantities of data which emerge.

2.3. SCIENTIFIC ARGUMENTS

2.3.1. *Improved signal-to-noise*
As already stated in the Introduction, the advantages of a statistical approach stem from the great reduction in statistical fluctuations. It is not the only way in which this can be achieved, however.

A first alternative is simply to run a single computation with much larger $N$. Suppose we compare a single calculation with $N$ particles with statistical results from $n$ simulations with $N/n$ particles each. Naively we may expect that the signal-to-noise of some measured quantity (such as the half-mass radius) will be comparable in the two sets of calculations. However, the single large calculation will have taken far longer, even if the $n$ small simulations were carried out sequentially, simply because the effort of a simulation increases more quickly than linearly with $N$. Thus the advantage of the statistical approach is efficiency (Smith 1977).

A second alternative is to resample a single simulation. The assumption here is that successive samples are approximately independent. Little has been done to test this assumption (Miller 1975), and in some cases it is clearly false. For example, core collapse takes place at different times in different simulations, and it follows that successive measures of the core radius in a single simulation will be biased by the time at which it collapses.

### 2.3.2. *Parameter dependence*

Very often the purpose of $N$-body simulations is not simply to provide an answer to a question about a single system, but to study the way in which the behaviour of the system changes as we change its initial or boundary conditions (e.g. the initial structure of the model). Nothing can be said about this, however, if one is restricted to a single simulation for each set of conditions (McMillan *et al.* 1990), for it is then possible that the changes observed are simply a result of statistical fluctuations and are not causally related to the changes in the imposed conditions.

### 2.3.3. *Incoherent Phenomena*

There are some important investigations for which a statistical approach is inappropriate. Incoherent phenomena (e.g. gravothermal oscillations) cannot be studied in this approach, at least not in the naive manner of combining data from different simulations at the same time. As another example, consider the implications of the fact, already mentioned, that different simulations end core collapse at different times. Though the averaged system will exhibit a collapse, it will not be as deep as that of any of the systems in the ensemble.

### 2.4. PREVIOUS STATISTICAL STUDIES OF $N$-BODY SYSTEMS

Numerous statistical studies of the three- and four-body problems have been reported in the literature. Such quantities as exchange cross sections could not, however, be obtained in any way other than by statistical study of large numbers of individual cases.

The earliest study in the spirit of this review was a set of twelve 5-body simulations carried out by Agekyan & Baranov (1969). Their aim was to measure the density distribution. This topic was one of several taken up subsequently by Smith (1977, 1979, 1985) who studied results from a small number of cases with $N$ in the range $8 \leq N \leq 64$. One aim of his investiga-tions was to determine empirically whether integration errors would have a noticeable effect on the statistical results. Heggie (1991) continued this theme with somewhat larger numbers of simulations and stars. A more systematic attack with statistical methods was initiated by Casertano (1985) and McMillan *et al.* (1988), whose main aim was to study the $N$-dependence of relaxation. This was also one of the early aims in the rather extensive studies by Giersz & Heggie, which will be summarised in more detail below. Finally, mention may be made of the recent study by Kroupa (1995) of clusters initially consisting entirely of binaries. He studied between 3 and 20 systems with $N = 400$ each, with a view to investigating the distribution of binaries in the field, and related problems.

2.5. THE STUDY BY GIERSZ & HEGGIE

The final goal of this investigation was to compare the results of Fokker-Planck and $N$-body simulations. The former was used by Chernoff & Weinberg (1990) in a landmark study of the evolution of globular clusters under the action of (i) mass loss by stellar evolution, (ii) two-body relaxation, and (iii) a steady tide. Because of the approximations used in the Fokker-Planck treatment it seemed worth while to check its results with $N$-body models, but the approach adopted was a gradual one, various dynamical processes being added one by one.

The first set of results analysed in detail had equal masses with no tide, and was studied in two papers: one (Giersz & Heggie 1994a; hereafter Paper I) dealing with evolution up to core collapse, and a second (Giersz & Heggie 1994b, = Paper II) dealing with core bounce and post-collapse evolution. Next came a study of systems with a mass spectrum, taken to be a power law (Giersz & Heggie 1995a, = Paper III). Later papers (Giersz & Heggie 1995b,c, = Papers IV, V) will add the effects of a steady tide (from which a small sample of results can be found in Giersz & Heggie 1993) and then mass loss from stellar evolution, some results from which are discussed below (§3).

In our work values of $N$ in the range $250 \leq N \leq 2000$ were used, the number of runs in each series varying from 16 (at the largest $N$) to about 90 (at smaller $N$). Note that, from a naive statistical point of view, 16 simulations with $N = 2000$ yield results of comparable statistical quality to a single run with $N = 32000$. In addition to the runs with either equal masses or a power-law mass function, some sets of runs of two-component models were studied and have been partially published elsewhere (Spurzem & Takahashi 1995).

Among the topics discussed in the course of these investigations have been the following.

TABLE 1. Some Homological Solutions for Cluster Evolution

| Source | Phase | BC | Model |
| --- | --- | --- | --- |
| Hénon 1961 | pc | tidal | Fokker-Planck |
| Hénon 1965 | pc | isolated | Fokker-Planck |
| Lynden-Bell & Eggleton 1980 | cc | power-law | gas |
| Inagaki & Lynden-Bell 1983 | early pc | power-law | gas |
| Goodman 1984, 1987 | pc | isolated | gas |
| Heggie & Stevenson 1988 | cc, early pc | power-law | Fokker-Planck |
| Louis 1990 | cc | power-law | anisotropic gas |
| Takahashi & Inagaki 1992 and Takahashi 1993 | cc, early pc | power-law | Fokker-Planck |

Explanation: BC, boundary condition; cc, core collapse; pc, post collapse.

- *The Coulomb logarithm* in the expression for the relaxation time. Usually taken as $\ln(\gamma N)$ with $\gamma = 0.4$ (Spitzer 1987), we found that a lower value of $\gamma \simeq 0.11$ fitted the data for core collapse better. Somewhat different values are indicated for post-collapse evolution, and smaller values for unequal masses (Papers I–III).
- *The escape rate* increases during core collapse, partly because of the change of structure, and partly because of the evolving anisotropy. It has also been measured for post-collapse evolution and for models with unequal masses and in a tidal field (Papers I–V).
- *Statistics of binaries*, including the maximum energies which they reach before ejection (Papers I–V).
- *Energy generation by binaries*, whose effectiveness appears to increase with $N$ in the range studied (Paper II).
- *Mass segregation*, which appears virtually to stop by the close of core collapse (Paper III).
- *Homological evolution*. Though such solutions seem very special (Table 1), it is remarkable how much of the evolution of even quite a complicated model can be understood in these terms (Fig.1). Homologous models also provide a means of encapsulating the very complicated behaviour of $N$-body models in quite simple approximate formulae (Papers I–V).
- *Effect of different types of tidal model:* it turns out that it makes little difference to the lifetime of a model if the tide is modelled as a simple cutoff (Paper IV).
- *Anisotropy*: as expected, in the presence of a tide it is difficult to produce anisotropy from isotropic initial conditions (Papers IV, V).
- *Scaling to real clusters*, which will be discussed in the next section.
- *Validation of Fokker-Planck and gas models.* One of the significant applications has been to check and calibrate these simplified models (Papers I, II; Giersz & Spurzem 1994). Essentially this brings up to date the older comparison of Aarseth *et al.* (1974). One major implication which deserves to be highlighted is that these comparisons (e.g. between the results of $N$-body and Fokker-Planck models) lend strong support to the Chandrasekhar-Spitzer model of two-body relaxation, with the cutoff radius at a value comparable with the system size, de-spite claims to the contrary (e.g. Gurzadyan & Savvidy 1986, Smith 1992).

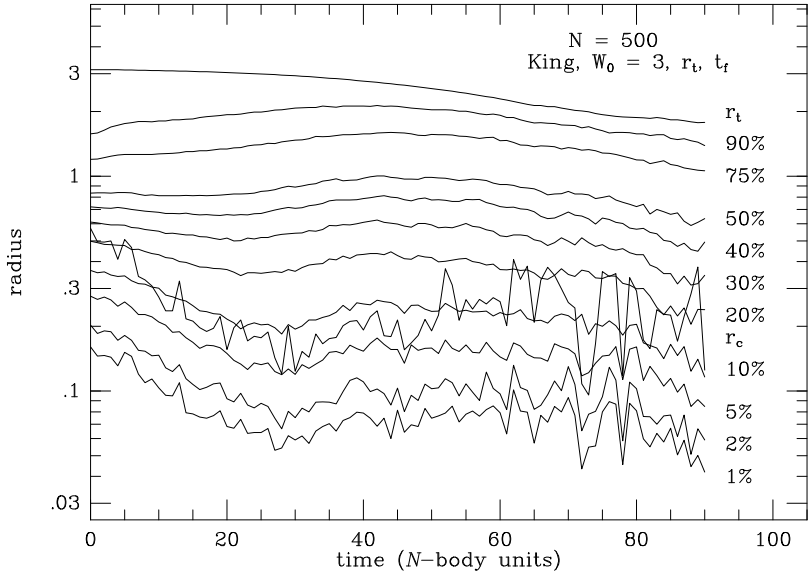

*Figure 1.* An $N$-body model exhibiting several types of nearly homologous behaviour (after Giersz & Heggie 1995b). Lagrangian radii corresponding to the stated mass fractions, and the core and tidal radii, are plotted against time in $N$-body units (Heggie & Mathieu 1986). The initial conditions are a 500-body King model with $W_0 = 3$, and a mass function with index 2.5 (Salpeter is 2.35) in a range such that $m_{max}/m_{min} = 37.5$ (cf. Chernoff & Weinberg 1990). The systems are tidally bound, but there is no mass loss from stellar evolution. Initially there were 56 cases, but the number of cases still running (determined by the condition $N > 25$) reduces rapidly between $t = 70$ and 90. This, together with the small number of very massive stars, accounts for the increasing noise in the innermost radii. Up to $t \simeq 30$ the core exhibits a nearly-homologous collapse. From then until $t \simeq 50$ there is almost self-similar post-collapse expansion as if the system were isolated. Thereafter the expansion is stopped by the contracting tidal radius, but the subsequent evolution is still nearly self-similar.

## 3. The Scaling Problem

One of the principal aims of $N$-body modelling is the understanding of globular clusters and their dynamical evolution. Even with modern special-purpose hardware, however, the values of $N$ for which reasonable numbers of simulations are feasible still fall short of the real numbers of stars in globular clusters. Therefore it is necessary to consider how the results scale with $N$. The essential problem is that evolution is driven by a variety of processes which scale in different ways with $N$ (McMillan 1993). In this section we illustrate this problem in connection with evolution driven only by two-body relaxation and mass-loss from stellar evolution in the presence of a galactic tide, with unequal masses.

This problem was considered by Chernoff & Weinberg (1990; hereafter CW) in their well known survey (see also Weinberg 1993). Consider as an example a cluster with initial mass $1.5 \times 10^5 M_\odot$, galactocentric distance 4kpc, initially a King model with scaled central potential $W_0 = 3$, and an initial mass function $dN/dm \propto m^{-2.5}$ in the range $0.4 M_\odot < m < 15 M_\odot$. We adopt the parameters for stellar mass loss and the galactic tide from CW.

The $N$-body results of Giersz & Heggie (1995c), with $N = 500$ and $1000$, showed several phases of evolution, which could be readily interpreted theoretically. Initially the cluster expanded, in response to the epoch of major mass loss by stellar evolution. This was followed by almost homologous contraction, as stars escaped across the tidal boundary. After 100 $N$-body units of time ($\simeq 1.7 \times 10^{10}$yr) there were the first signs of core collapse. Unfortunately, CW's Fokker-Planck result was quite different: they found that the cluster dissolved after $2.8 \times 10^8$yr!

The problem here is one of scaling, as Table 2 illustrates. The conversion between $N$-body time units and astrophysical units which was assumed above was guided by the correct scaling of the relaxation time, and is denoted $N$-body-A in the table. Because the crossing time scales differently with $N$, however, this means that the crossing time in the $N$-body model does not then scale to the correct value in years. As can be seen in the Table, this implies that mass-loss is impulsive in the $N$-body model, i.e. it takes place on a time scale short compared with the crossing time, whereas it is adiabatic in the "real" cluster. At first sight it is therefore puzzling that it is the $N$-body model which survives better: for an isolated cluster, simple virial estimates show that arbitrary amounts of mass may be lost adiabatically, but only 50% impulsively. It is possible that this distinction

88                           DOUGLAS C. HEGGIE

TABLE 2. Scaling of $N$-body models

| System    | $t_{cr}$          | $t_{se}$      | $t_{rh}$            | $N$      |
|-----------|-------------------|---------------|---------------------|----------|
| Cluster   | $5 \times 10^6$yr | $10^8$yr      | $3 \times 10^9$yr   | $10^5$   |
| $N$-body-A | 3                | 0.5           | 14                  | 500      |
| $N$-body-B | 3                | 60            | $\gg 250$           | 16000    |

Explanation of symbols and units: $t_{cr}$, crossing time; $t_{se}$, time of mass loss by $4M_\odot$-star; $t_{rh}$, half-mass relaxation time; $N$, number of stars; for the $N$-body models the unit of time is standard (Heggie & Mathieu 1986).

between the effects of adiabatic and impulsive mass loss is altered in the presence of a tide.

In the real cluster the three time scales are well separated, and it is clearly necessary to use an $N$-body simulation with the same ordering of time scales. This is difficult without increasing $N$ substantially. Nevertheless this is now quite feasible, thanks to the development of GRAPE special-purpose hardware (Ebisuzaki et al. 1993; papers by Makino and Taiji, these proceedings.) Using a low-precision GRAPE-3AF at Edinburgh it was possible to increase $N$ large enough so that the crossing- and stellar-evolution times scaled consistently, and the relaxation time was large enough not to affect the evolution. This model is denoted $N$-body-B in the table. (Note that, in the Fokker-Planck model, the evolution is over in about $0.1t_{rh}$.) The result (Fukushige & Heggie 1995) is that the cluster dissipates after about $9 \times 10^8$yr, i.e. longer than the lifetime obtained by CW, but by a factor of only about 3. The residual disagreement is thought to be due to the fact that the disruption takes place on a time scale which is not sufficiently large compared with the crossing time. Under these circumstances one of the assumptions underlying the use of the orbit-averaged Fokker-Planck equation, which were stated carefully and explicitly by CW, is not satisfied.

## 4. Conclusions and Reflections

The following conclusions are justified by the arguments of the foregoing sections.

First, there are several reasons for arguing that not just one but several $N$-body simulations should be carried out with the same parameters, differing only in the random numbers used in generating initial conditions. In particular, there are significant scientific benefits in combining results from large numbers of simulations.

The second set of conclusions concerns the problem of scaling results from simulations to real clusters, when the simulations are carried out with modest numbers of particles (i.e. small compared with the number of stars in a real cluster). In particular, we have shown that, while mass-loss by stellar evolution is important, it is necessary to scale the results so that the time scale for mass loss and the crossing time both scale consistently to the relevant astrophysical values. Thereafter, when mass loss becomes relatively unimportant, the scaling would have to alter so that the relaxation time scales correctly. Similar problems are presented by the modelling of disk shocking, whose inclusion in fully-fledged $N$-body simulations has not yet been attempted. In both cases, there may be sets of parameters for which no meaningful $N$-body simulations can be attempted with modest values of $N$ (i.e. thousands).

A third and rather minor conclusion is that some caution should be exercised in using the orbit-averaged Fokker-Planck equation for the modelling of very young clusters. More generally, it is well known that such models involve several simplifying assumptions and depend heavily on our theoretical understanding of the dynamical processes which drive the evolution of clusters. By contrast it is sometimes asserted that $N$-body models are essentially "assumption-free". It is worth pointing out, however, that $N$-body models also require considerable theoretical input, as we have seen, if it is desired to scale the results intelligently to real star clusters.

## Acknowledgements

The author thanks D. Sugimoto and his colleagues in the GRAPE group for their enthusiastic help in setting up GRAPE hardware in the UK. The author's own recent visits to Tokyo have been supported by travel grants under a cooperative scheme of the Royal Society, the Japanese Society for the Promotion of Science, and the British Council (Tokyo).